**Physical inversion of the full IASI spectra: further validation and inter-comparison of $O_3$ and OCS products.**

G. Liuzzi[a], G. Masiello[a], C. Serio[a,*], S. Venafra[a] and C. Camy-Peyret[b]

[a]*Scuola di Ingegneria, Università della Basilicata, Potenza, Italy*
[b]*Institut Pierre-Simon Laplace (IPSL), UPMC/UVSQ, Paris, France*

*Corresponding author
e-mail address: carmine.serio@unibas.it

*Abstract.* Results from the physical, simultaneous, inversion of the full IASI (Infrared Atmospheric Sounder Interferometer) spectra applied to the retrieval of surface temperature and emissivity, temperature, water vapour, ozone and HDO vertical profiles, and gas column abundance of $CO_2$, CO, $CH_4$, $SO_2$, $N_2O$, $HNO_3$, $NH_3$, OCS and $CF_4$ have been updated with respect to the validation of OCS and $O_3$ products. A case study has been set up and analysed, which consists of a two years long record of IASI spectra over sea surface in the Pacific Ocean close to the Mauna Loa (Hawaii) validation station. IASI $O_3$ products have been intercompared with OMI (Ozone Monitoring Instrument) $O_3$ retrievals, which, in contrast to IASI that uses the thermal infrared, are based on observation in the near ultra violet region of the electromagnetic spectrum. IASI OCS retrievals are compared to *in situ* flask observations. The IASI vs OMI comparison shows an excellent agreement with a correlation of 0.97 and a relative bias of $\approx$ 1 Dobson Unit. For the case of OCS, we have found a good correlation of 0.75 for night time soundings. IASI captures the OCS seasonal cycle, although a relatively large negative bias of $\approx$ -74 pptv is seen, which can be explained with a large spectroscopic error for OCS in the HITRAN 2008 version.

1. Introduction

In a recent paper [1] the authors have presented a simultaneous physical retrieval approach for IASI (Infrared Atmospheric Sounder Interferometer) [2], which uses all available observations on the spectral coverage extending from 645 cm$^{-1}$ to 2760 cm$^{-1}$. The retrieval scheme has the twofold objective of a) performing a spectral residual analysis aiming at addressing quality and accuracy of spectroscopy and b) providing the simultaneous retrieval of all geophysical parameters, which contribute to the formation of the IASI spectra. These parameters include surface temperature and emissivity, temperature, water vapour, ozone and HDO vertical profiles, and gas column abundance of $CO_2$, CO, $CH_4$, $SO_2$, $N_2O$, $HNO_3$, $NH_3$, OCS and $CF_4$. The retrieval approach was exemplified in [1] based on a case study, which consisted of a two year long record of IASI spectra over sea surface in the Pacific Ocean close to the Mauna Loa (Hawaii) validation station.

In [1] the authors were mostly focused on the validation of the main green-house cases, namely $CO_2$, $CH_4$ and $N_2O$ and other gases of climatological and air quality concern: $O_3$ and CO. Because of lack of validation data, trace species were only introduced for completeness of the analysis and for a comprehensive analysis of spectral residuals.

The present short communication is complementing the information provided in [1] focusing on two interesting aspects. The first is the comparison of IASI ozone retrieval vs that produced from OMI (Ozone Monitoring Instrument) observations. OMI flies on National NASA's Aura satellite and is part of the Earth Observation System (EOS). In contrast to IASI, which mainly retrieves ozone from $O_3$ absorption band at 9.6 μm, OMI measurements of ozone columns are based on observations within the Ultra Violet (UV) band 300 nm to 340 nm. This comparison is interesting in that it provides an indirect check of the consistency of ozone spectroscopy between the infrared and UV.

The second one is the validation of IASI OCS (carbonyl sulfide) retrieval with *in situ* observations at the Mauna Loa and Cape Kumukahi stations, Hawaii, USA. The importance of carbonyl sulfide in the study of ecosystems has clearly emerged in recent studies [3,4]. OCS is the most abundant sulfur-containing trace gas in the atmosphere, and carries a significant part of sulfur in the stratospheric aerosol [5]. Major sources of OCS are natural, and that among them oceans, soils and volcanic eruptions play a dominant role. Otherwise, anthropogenic sources are recognized to be of second order: the most important of them are biomass burning and industrial activities [6]. The main sink of OCS has been identified in the vegetation uptake, whose magnitude is also influenced by seasonal trends in vegetation growth. Conversely, in the stratosphere photochemical loss is the prominent process that remove OCS from the atmosphere. Moreover, OCS has recently emerged as a putative proxy for the photosynthetic uptake of $CO_2$, because OCS and $CO_2$ have the same diffusion pathway into leaves [7], and OCS hydration reaction in this process is irreversible. The quantification of OCS concentration on forest canopy is also of primary interest to obtain indications about the direct impact of photosynthetic processes on the $CO_2$ seasonal and inter-annual trends. Therefore, the capability of IASI to retrieve the seasonal cycle of OCS is an important asset, although in this paper we limit the discussion to the ocean case.

This paper is organized as follow. Section 2 summarizes data and methods. Section 3 will show the results and section 4 will be devoted to conclusions.

2. Data and methods.

IASI and ancillary data have been discussed at a length in [1]. Here we limit the discussion to the new data used for the inter-comparison/validation of $O_3$ and OCS.

The target area of interest is shown in Fig. 1. For this area, OMI $O_3$ data, daily values, (level 3 products, version OMSO2e) were acquired for the whole period Jan. 2014 to Dec. 2015. The native format of the data consists of gridded column amount of $O_3$ (Dobson Units, DU) available on 0.25 degrees Lat/Lon grid. The data were downloaded from the website http://disc.sci.gsfc.nasa.gov/Aura/data-holdings/OMI/omso2e_v003.shtml.

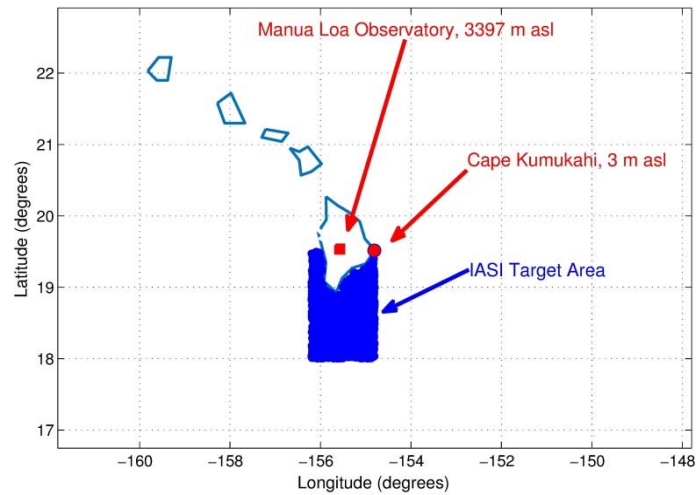

Fig. 1 Case study target area showing the region for which IASI and OMI $O_3$ data where acquired for the period 2014-2015. The figure also shows the location of Mauna Loa and Cape Kumukahi station.

OCS flask weekly data were collected at both stations of Mauna Loa (shortname MLO, 19.5362°N, 155.5763°W, 3397 m asl) and of Cape Kumukahi, (shortname KUM, 19.516°N, 154.811°W, 3 m asl) during the period 2014-2015. MLO and KUM observatories in Hawaii are part of the National Oceanic and Atmospheric Administration (NOAA), of the Earth System Research Laboratory (ESRL) and of the Global Monitoring Division (GMD). OCS data were downloaded from the website http://www.esrl.noaa.gov/gmd/hats/gases/OCS.html (see also [3] for a detailed description of these data). KUM is a station at sea level, whereas MLO is at an altitude of 3397 m and, therefore, the data here discussed are representative of the free troposphere, to which IASI is most sensitive. The comparison of MLO to KUM is indicative of how much the seasonal cycle amplitude is attenuated with altitude.

The retrieval methodology has been described in detail in [1]. We outline here only the main elements of the procedure, in which the full IASI spectrum (8461 channels) is used to simultaneously retrieve, the atmospheric profiles of temperature ($T$), water vapour ($Q$), ozone ($O$), and HDO, the surface temperature ($T_s$), and the columnar amount of the following gases: $CO_2$, $N_2O$, $CO$, $CH_4$, $SO_2$, $HNO_3$, $NH_3$, OCS and $CF_4$. For land, the methodology retrieves also surface emissivity ($\varepsilon$). The emissivity spectrum (at IASI sampling of 0.25 cm$^{-1}$) is parameterized through a truncated Principal Component (PC) transform. The first 20 PC scores are retained, hence retrieved. For sea surface, emissivity is not retrieved and the Masuda model [8] is fixed in the retrieval.

An important improvement over the version used in [1] is the dimensionality reduction of the data space through *random projections* (RP) [9]. This transform allows us to reduce the dimensionality of IASI observations from 8461 channels to 1200 RPs. In addition to the dimensionality reduction, the transform is allowing a consistent and unified treatment of both systematic and random error sources affecting the infrared observations.

The inversion algorithm is based on the integration between a fully analytical, monochromatic radiative transfer model, and an Optimal Estimation module [10]. The radiative transfer model is called σ-IASI [11,12,13,14] that covers the spectral range 10 to 3000 cm$^{-1}$. The σ-IASI code is a fast model, which uses a pre-computed look-up table for the optical depth calculation. The look-up table is derived from LBLRTM (Line-by-Line Radiative Transfer Model, [15]), version 12.2, which adopts the HITRAN 2008 line compilation [16]. The forward module computes the radiance, discretizing the radiative transfer equation on a fixed grid made up by 60 pressure layers, spanning the atmosphere from 1050 to 0.005 hPa. The forward model computes spectral radiances and analytical Jacobians with respect to surface temperature, emissivity and any atmospheric parameter, including the mixing ratio profile of the aforementioned gases. Forward model IASI radiances are obtained through convolution with the IASI Instrumental Spectral Response Function (ISRF).

The inversion scheme, called δ-IASI [17] implements an iterative algorithm for the optimal estimation of the thermodynamic state of the atmosphere and its composition. The model δ-IASI performs the mathematical inversion of the whole IASI radiance spectrum (8461 channels) to simultaneously retrieve the state vector ($\varepsilon$, $T_s$, $T$, $Q$, $O$, HDO, $CO_2$, $N_2O$, $CO$, $CH_4$, $SO_2$, $HNO_3$, $NH_3$, OCS and $CF_4$). One important aspect on the retrieval scheme, which is relevant to the estimation of trace gases, is the use of background and first guess state vectors which are in the linear region of the radiative transfer equation. Towards this objective we use the ECMWF (European Centre for Medium range Weather

Forecasts) analysis for $(T, Q, O)$ collocated in space and time with the IASI soundings (see [1] for details). The use of ECMWF $(T, Q, O)$ analysis is a strong constraint for the inverse scheme, which allows us to relax that for trace gases. In effect, these are normally adjusted from their climatological first guess to their final values in one iteration.

## 3. Results

*3.1 Ozone*

For the case of ozone the inversion schemes retrieves the profile over a pressure grid of 60 layers. The total column of ozone is obtained by proper integration on the pressure grid. For details on the ozone background (state vector and covariance matrix) the interested reader is referred to [1]. Figure 2 shows the comparison (monthly mean) between IASI-$O_3$ and OMI-$O_3$. For completeness, *in situ* MLO observations are also shown. Working in the infrared IASI can provide day and night time observations. The correlation of IASI and OMI with MLO observations is 0.97 and 0.98 respectively.

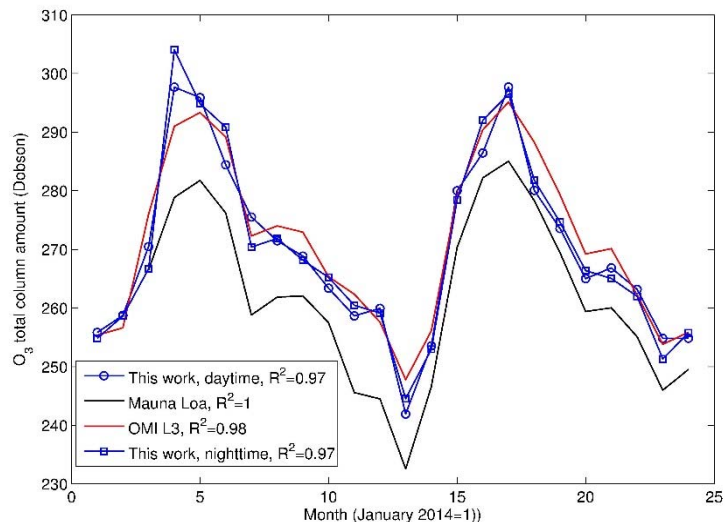

Fig. 2 IASI Retrieved monthly column amount (DU) for ozone and comparison with OMI and *in situ* data. IASI retrievals and OMI data have been spatially averaged over the target area shown in Fig. 1.

Daytime and night-time IASI ozone retrievals do not show any systematic or mutual inconsistency, which shows how the IASI retrieval system is capable to work day and night with the same accuracy. The correlation between IASI daytime and OMI is 0.97 and the mean difference between the two is 1.44 DU. The correlation of IASI night-time is again 0.97 but now the mean difference is slightly negative, -1.12 DU. It is also seen that *in situ* data are systematically smaller than OMI and IASI satellite observations. In fact, the station is located at 3397 m and therefore MLO observations are not sensitive to the lowermost tropospheric ozone. The ozone layer from sea level to MLO has been assessed with ozonesonde observations, as in [18] and depending on the month the lower troposphere ozone layer is 7-10 DU thick.

A similar exercise was performed in 2005 [19] between IMG (interferometric Monitoring of Greenhouse gases, a precursor of IASI) and TOMS (Total Ozone Mapping Spectrometer) flying on the Japanese platform ADEOS (Advanced Earth Observing Satellite). At the time the correlation was found to range in between 0.6-0.7, which exemplifies the progress in remote sensing technology and methods, but also the improvement of ozone spectroscopy (see e.g. [20-22]).

3.2 OCS

For each trace gas, the retrieval scheme retrieves a *scaling factor*, which multiplies the *a priori* or background vertical profile fixed according to the AFGL (Air Force Geophysical Laboratory) climatology [23]. This is true also for OCS, which we will discuss in terms of total column amount. For trace gases the methodology has been validated for $CO_2$, $N_2O$, CO and $CH_4$ by comparison with *in situ* observations and satellite equivalent products in [1].

OCS spectral signatures are well detectable in IASI radiances in the spectral range 2030-2085 cm$^{-1}$. Since we use Optimal Estimation, the retrieval algorithm needs a suitable background state. For trace species, the background is fixed by climatology. For the specific case of OCS the background profile, $q_a(p)$ (with the $p$ the pressure coordinate) is shown in Fig. 3 and has a total column amount of 510 pptv. Within the retrieval scheme, the OCS profile, $q(p)$ is parameterized with a single scaling factor, that is $q(p)=(1+f)q_a(p)$ and the retrieved parameter is the scaling factor $f$. Let $\hat{f}$ be the retrieved scaling factor, the retrieved total column amount, say $\hat{Q}$ is given by

$$\hat{Q} = (1 + \hat{f})Q_o \qquad (1)$$

with the background column amount $Q_o = 510$ pptv. The background we use for $f$ is $f_a = 0$ and the associated variance is $\sigma_{f_a}^2 = 2$. This choice is consistent to $dof = 1$, as will be shown in the following.

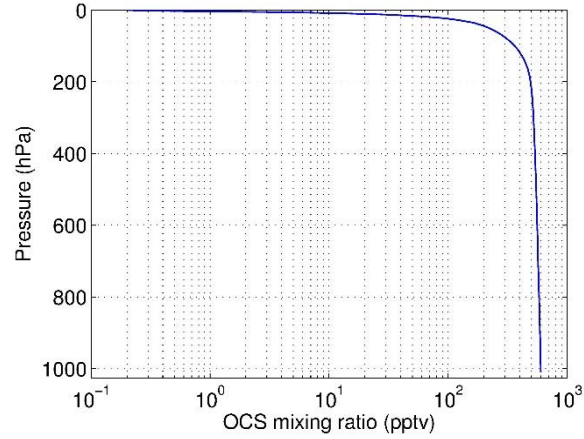

Fig. 3 OCS mixing ratio profile used as background in the present analysis.

Because, the column amount is parameterized with a single scaling factor, OCS contributes to the state vector for one degree of freedom or $dof$. The actual $dof$ of OCS retrieval depends on the variance of the background $\sigma_{f_a}^2$. This dependence can be calculated in simulation based on a linear error analysis and is shown in Fig. 4 for a typical tropical air mass. The larger $\sigma_{f_a}^2$, the larger the $dof$. With $dof = 1$, the retrieval loses its dependence on the background and will be determined by the data alone. It can also be shown (e.g. [17]) that in the limit $\sigma_{f_a}^2 \to \infty$ the bias affecting the solution tends to zero. However, this may occur at the expense of the retrieval precision, which is the *a posteriori* retrieval standard deviation. Thus, we have to trade-off between precision and bias. Figure 4 shows the retrieval precision as a function of the *a priori* or background variance $\sigma_{f_a}^2$. Noting that

$$\text{var}(\hat{Q}) = \text{var}(\hat{f})Q_o^2 \qquad (2)$$

(where var stand for variance) the fractional precision of the retrieval can be written as

$$\sqrt{\frac{\text{var}(\hat{Q})}{Q_o^2}} = \sqrt{\text{var}(\hat{f})} = \sigma_{\hat{f}} \qquad (3)$$

that is the standard deviation (shown in Fig. 4) of the scaling factor retrieval, $\hat{f}$.

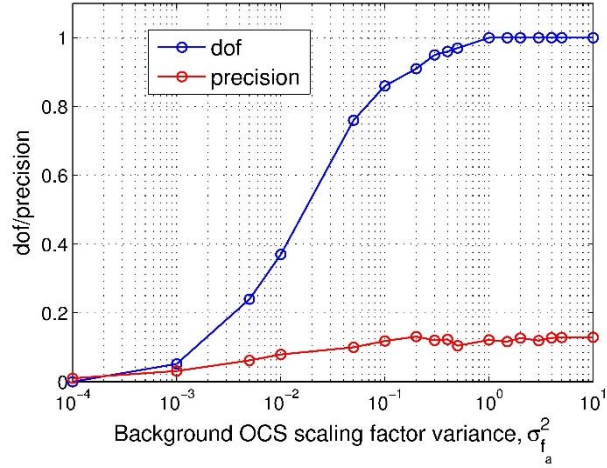

Fig. 4. Retrieval *dof* and precision as a function of the background OCS scaling factor variance.

From Fig. 4 it is seen that the *dof* saturates to 1 for $\sigma_{f_a}^2 \geq 1$. After an expected linear behaviour (this is because for *dof* close to zero, the variance retrieval has to be equal to background variance) the precision tends to a constant value of about 12% (some fluctuations of the precision for large $\sigma_{f_a}^2$ are expected because we use *random projections* [8] and therefore the basis changes for each single run). It is interesting to see that the precision does not diverge, which means that the unconstrained retrieval for OCS is possible. This condition is equivalent to an unconstrained Least Squares minimization. The good conditioning is largely due to our parametric approach, which represents the OCS profile with a scaling factor. It should be stressed that for two generic OCS profiles, say $q(p)$, $q_a(p)$, of whatever shape, their integral or column amount can be made equal just by defining an appropriate scaling factor, $f$ such that

$$(1 + f) \int_{p_g}^{0} q_a(p) dp = \int_{p_g}^{0} q(p) dp \qquad (4)$$

with $p_g$ the ground pressure. Apparently, this approach has been also used in a concurrent paper [24], which uses a background standard deviation of 2 (200%). They report a retrieval precision of about 12%, which is consistent with the present analysis (see Fig. 4).

An alternative approach [10] would be to use a non-parametric estimation, where a vector-valued profile **q** of size *N* is first retrieved and then the column amount is obtained as

$$\hat{Q} = \boldsymbol{h}^t \hat{\boldsymbol{q}}; \quad \text{variance}(\hat{Q}) = \boldsymbol{h}^t \boldsymbol{S}_{\hat{\boldsymbol{q}}} \boldsymbol{h} \qquad (5)$$

with $\boldsymbol{S}_{\hat{q}}$ the retrieval covariance matrix and **h** the averaging operator (e.g. see [1]). This approach needs a proper background covariance matrix, $\boldsymbol{S}_{q_a}$. The retrieval independence from the background is reached provided the variances $S(i,i)_{q_a} \to \infty; i = 1, \ldots, N$. However, in this limit, the related unconstrained Least Squares problem to get $\hat{\boldsymbol{q}}$ does not have solution or equivalently the retrieval covariance matrix becomes infinity. In practice, if information content is limited to $dof \leq 1$, the above non-parametric approach has to rely on the background to stabilize the inverse process. Even in this case, the retrieval precision could become so poor that one needs to perform consistent spatial and time averaging.

Apparently, the approach of Eq. (5) was used for OCS in [25]. However, they had to average the retrieval on a 10° Lat/Lon grid box and then to further average the results on a multi-year sample of 5 years.

In this paper we use, the approach of Eq. (1) with $\sigma_{f_a}^2 = 2$, which correspond to $dof = 1$ and a retrieval precision for a single IASI field view was ≈ 12%. In passing, we note that the results shown in [1] for OCS (see Fig. 26 in that paper), were obtained with a background standard deviation of 0.1, therefore $dof \approx 0.4$ and the retrieval was mostly determined by the background.

Finally, in the present analysis, to avoid outliers caused by non-linearity in the radiative transfer equation, only retrievals that converged at the first iteration were considered.

The results are shown in Fig. 5 along with a comparison with MLO and KUM *in situ* observations. Only night-time IASI retrieval have been used to avoid problem with solar contamination and non-LTE effects in IASI band 3. This allows a better interpretation of the bias between *in situ* and IASI observations.

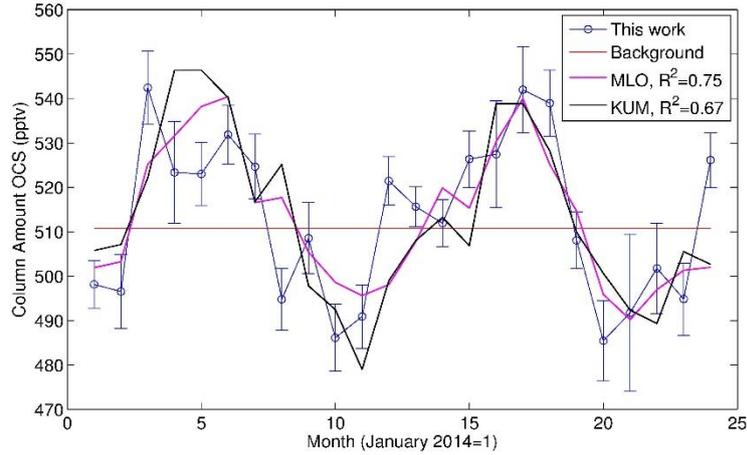

Fig. 5. IASI retrieved monthly column amount (pptv) for OCS and comparison with MLO and KUM *in situ* data. IASI retrievals have been spatially averaged over the target area shown in Fig. 1. The error bar represents the error in the mean (standard deviation of soundings in the target area for the given months divided by the square root of the number of observations). $R^2$ in figure is the linear correlation coefficient of *in situ* observations against IASI retrieval. A bias of -74 pptv has been subtracted from the IASI OCS retrieval.

From Fig. 5, it is seen that the correlation with KUM data is 0.67, which increase to 0.75 in the case of MLO. This is not surprising since IASI observations over the ocean are not sensitive to the gas concentration close to the surface. In effect, an analysis of the radiance Jacobian for OCS (see Fig. 6) shows that the IASI is mostly sensitive to the middle-upper troposphere. The sensitivity tends to zero close to the surface. In addition, it is known [3] that the OCS cycle tends to decrease its amplitude with altitude.

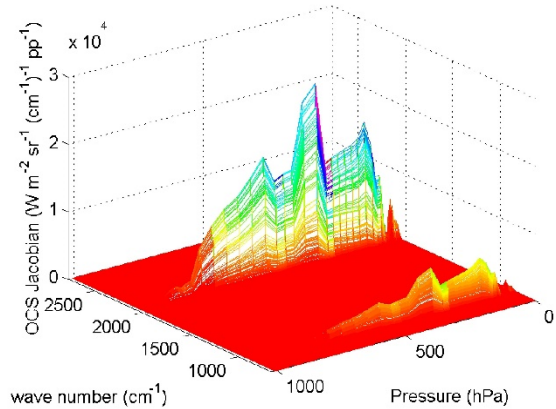

Fig. 6. OCS Jacobian for the IASI range 645 to 2760 cm$^{-1}$.

It is interesting to note that IASI can resolve the full amplitude of the seasonal cycle as seen at MLO: the standard deviation of the OCS time series is 18 pptv for IASI and 15.5 pptv for MLO. Comparing these results to those shown in [19], we see that [19] is resolving only about 40% of the cycle amplitude, which is likely resulting from the retrieval dependence on the background. In contrast, with the present work approach, IASI seems to resolve the full cycle.

It is also important to stress that IASI OCS is biased with respect to MLO, the bias (IASI-MLO) is $\approx$-74 pptv. A negative consistent bias was found also in [22]. Although the amount of bias was smaller, most likely because the retrieval dependence on the climatological background, whose magnitude is close to MLO OCS average.

The source of this bias cannot be due to interfering factors because we retrieve all parameters. It cannot be the effect of solar contamination and non-LTE effects, because we use night time soundings alone. The only relevant source remaining is spectroscopy. Our forward model uses HITRAN 2008 [13] and we know [26] that in HITRAN 2008 OCS compilation was suffering from various software errors and as a consequence some of the air- and self-broadening half widths for all isotopologues carbonyl sulfide were in error by as much as 50%. In effect a look at the spectral residuals (Fig. 7) in the OCS spectral region 2020 to 2085 cm$^{-1}$ shows a consistent bias (negative Obs-Calc) that does not go to zero.

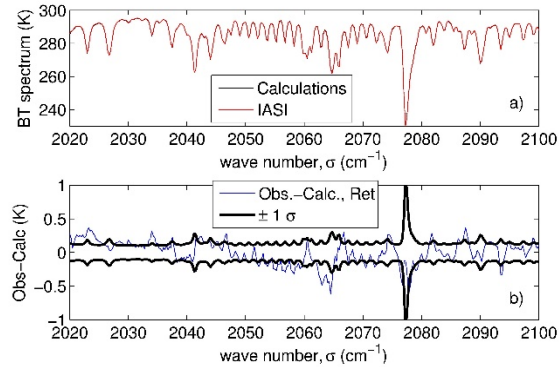

Fig. 7. Observed and Calculated IASI spectra (a) and related Obs-Calc residuals (b). Spectra have been averaged over the period 2014-2015 (2276 IASI soundings). The ± 1 σ error bar in panel b) is the IASI radiometric noise, shown to provide a reference for comparison with the bias amplitude.

## 4. Conclusions

An update of the results shown in [1] has been presented, which mostly focused on $O_3$ and OCS. Satellite based infrared sounding of ozone has reached a high level of maturity in terms of spectroscopy and of retrieval algorithm to compare well with UV satellite observations.

We have also shown that IASI is sensitive to the OCS seasonal cycle over ocean, although a large bias is still affecting the retrieval mostly because of a well-known OCS spectroscopy inconsistency in the HITRAN 2008 compilation. We think that for land OCS should even be an easier target for IASI because of the larger thermal contrast.

Finally, the analysis has shown that trace species can be retrieved from IASI without heavy constraint, provided the $(T, Q, O)$ background and first guess state vectors are as good as to linearize the radiative transfer equation.

**Acknowledgements**

Observations recorded at Mauna Loa and Cape Kumukahi Observatories in Hawaii have been downloaded from the web page http://www.esrl.noaa.gov/gmd/hats/gases/OCS.html. These observatories are part of the National Oceanic and Atmospheric Administration (NOAA), the Earth System Research Laboratory (ESRL) and the Global Monitoring Division (GMD). We give credit to the National Oceanic and Atmospheric Administration (NOAA)/Global Monitoring Division (GMD) in Boulder as a source of the data. IASI has been developed and built under the responsibility of the Centre National d'Etudes Spatiales (CNES, France). It is flown onboard the Metop satellites as part of the EUMETSAT Polar System. The IASI data L1C and L2 are received through the EUMETCast near real time data distribution service.